# An Interactive Platform for Unified Assessment of Drug-Drug Interactions Using Descriptive and Pharmacokinetic Data

*Nadezhda Diadkina*

*University of Lisbon, Lisbon, Portugal*

**Abstract**

Drug-drug interactions (DDIs) are a major concern in polypharmacy. Public databases often provide only qualitative descriptions without pharmacokinetic context. We present an interactive web tool that integrates 191,541 descriptive DDI records from DrugBank with 3,779 AUC-based interactions from the PK-DDIP dataset, extracted from FDA-approved drug labeling (DailyMed) using natural language processing and manual verification. Using multi-step name harmonization (exact, fuzzy, synonym expansion, manual curation), 1,803 interaction pairs were unified across 639 overlapping compounds. The platform is built with Streamlit and offers smart drug search, color-coded output, and real-time response. It enables users to explore both descriptive and pharmacokinetic data, supporting research and education in clinical pharmacology and drug safety. This research prototype is not intended for clinical decision-making.

## 1. Introduction

Modern therapy frequently involves concurrent use of multiple medications, increasing the risk of drug–drug interactions (DDIs) [1]. For example, more than one-third of older adults in the United States regularly take five or more medications, and about 15% of these individuals are at risk of a serious DDI [1]. Likewise, a study in community pharmacies found that approximately 39% of reviewed prescriptions contained at least one potential DDI [2]. A recent meta-analysis confirms the high prevalence of potential DDIs among older adults and highlights the variability in reported rates due to inconsistent identification methods [3], [4]. Such interactions can result in reduced therapeutic efficacy or dangerous adverse effects, including adverse drug reactions and even hospitalizations [2]. Unrecognized interactions have led to serious adverse outcomes; e.g., astemizole-related arrhythmias with CYP3A4 inhibitors contributed to market withdrawal [5], [6], [7]. These examples underscore the critical importance of identifying and managing DDIs to ensure medication safety. Mechanistically, DDIs are pharmacodynamic or pharmacokinetic (PK). Pharmacodynamic interactions arise when drugs affect the same targets or pathways, producing additive, synergistic, or antagonistic effects [8]. PK interactions occur when one drug alters absorption, distribution, metabolism, or excretion of another, changing systemic exposure [8]. For example, individual nonsteroidal anti-inflammatory drugs (NSAIDs) (described for specific molecules) can inhibit the metabolism of the anticoagulant warfarin, causing elevated warfarin levels and a significantly increased risk of bleeding [1]. Understanding metabolic pathways and clearance is therefore essential for anticipating harmful PK DDIs.

Despite extensive efforts to document drug interactions, many potential DDIs remain unknown or poorly characterized. With every new approval, the combinatorial space of drug pairs explodes, and it is infeasible to empirically test all combinations [1], [5], [9], [10]. Computational (*in silico*) methods can screen large numbers of pairs, but most models predict occurrence rather than magnitude or practical management strategies [5].

Alongside prediction tools, numerous knowledge-based resources support DDI checking (DrugBank, Drugs.com, Lexicomp, Medscape, DDInter, etc.) [8]. These are widely used in practice, but no single database is comprehensive. Different sources often disagree on the presence or severity of the same interaction, prompting recommendations to consult multiple databases [2]. For example, one comparative study found that the Medscape interaction checker identified more potential DDIs in patient prescriptions than either Drugs.com or Lexicomp did, and the severity ratings for the same interactions often differed between these databases [2]. Moreover, qualitative resources typically lack integrated PK context (enzymes/transporters, effect size), which impedes rapid interpretation at the point of need [5].

To address these gaps, we developed a tool that enables rapid DDI checks while integrating relevant PK information from open sources. The goal of this platform is to combine the breadth of existing DDI knowledge with the depth of pharmacokinetic details – allowing users not only to find out if two drugs interact, but also to understand how and why. The system aggregates interaction text from established databases and supplements it with PK annotations. It does not generate novel interaction data; rather, it organizes reported information into an accessible, unified interface. We outline the data pipeline and application, and compare its capabilities with existing resources.

## 1.1. Existing Drug Interaction Databases

A variety of drug interaction databases have been developed over the years. Qualitative compendia remain foundational in clinical settings. DrugBank provides comprehensive (over 30,000 drug entries), literature-derived narratives with severity classifications but generally lacks quantitative effect sizes. Commercial systems (Lexicomp, Micromedex) offer real-time screening with clinical guidance. These systems are valuable for alerting clinicians to potential risks, but they tend to be conservative in assessments – a design choice that contributes to high false-positive alert rates and "alert fatigue" in practice [11]. This phenomenon has been documented across multiple CPOE/CDS implementations, with high override rates and recommendations to minimize interruptive alerts for lower-risk DDIs [9]. Experts advise cross-referencing multiple sources for completeness [11], [12].

Consumer-oriented tools (e.g., Drugs.com, Medscape) broaden access but prioritize breadth over mechanistic depth and quantitation, which limits clinical utility in complex cases. For instance, one analysis of psychotropic drug interactions found that Drugs.com flagged the most interactions overall, whereas Lexicomp (a professional tool) covered a broader range of prescribed medications and provided more actionable clinical recommendations [13]. Quantitative/PK-focused resources (e.g., PK-DB; FDA/EMA clinical pharmacology reviews; subscription DIDB) provide AUC or Cmax changes and mechanistic evidence, but are fragmented and often not optimized for point-of-care use. Regulatory guidance explicitly defines study expectations and interpretation thresholds for CYP/transporter-mediated DDIs [9], while large curated corpora underpin subscription databases used in research and labeling.

Another important repository is the University of Washington Drug Interaction Database (DIDB), a subscription academic database compiling peer-reviewed studies of DDIs, especially those mediated by cytochrome P450 enzymes.

Recently, integrated and open-access initiatives have sought to combine the strengths of multiple sources. DDInter aggregates interaction information across literature and labels with mechanism and risk annotations [14], [15], but, like most amalgamated resources, remains largely qualitative. Recent harmonization efforts (ICH M12) aim to standardize the design, analysis, and reporting of DDI studies across regions, facilitating consistent integration into knowledge bases [16].

In summary, existing DDI databases vary widely in content and focus, ranging from text-based compendia to pharmacokinetic data repositories and specialized tools. No single resource is all-encompassing.

## 2. Methods

*2.1. Data Sources and Dataset Structure*

**Descriptive Drug-Drug Interaction Data**

The descriptive DDI dataset was extracted from DrugBank version 5.1.8, a comprehensive pharmaceutical knowledge base containing detailed information on drugs and their documented interactions [17], [18]. It comprises 191,541 unique interaction records among 1,701 distinct compounds. Each record includes drug1, drug2, and a narrative description of mechanism/clinical significance. These descriptions were derived from peer-reviewed literature, regulatory submissions, and expert clinical knowledge, providing a comprehensive foundation for qualitative DDI assessment.

**Pharmacokinetic Drug-Drug Interaction Data**

The PK dataset is PK-DDIP (Jang et al., 2022), containing AUC fold-change values extracted from FDA-approved product labeling (DailyMed) using standardized NLP plus manual verification. These values originate from information reported in product labeling, which may be based on dedicated clinical DDI studies, modeling, or other regulatory sources [5], [19], [20]. It includes 3,779 drug-drug interaction pairs among 741 unique drugs, each with a reported AUC fold-change for systemic exposure when co-administered versus alone. For coverage analyses we treat drug pairs as unordered (A–B ≡ B–A), yielding 3,092 unique pairs because directionality in PK-DDIP often reflects perpetrator/victim roles.

Each record in the PK-DDIP dataset includes the perpetrator drug name (drug1), victim drug name (drug2), and the corresponding AUC fold-change value (auc_fc). In the PK-DDIP dataset, the AUC fold-change represents the ratio of reported systemic exposure (AUC) for a victim drug when co-administered with a perpetrator drug, compared with administration alone, as documented in FDA labeling. Values greater than 1.0 indicate a higher reported exposure, values less than 1.0 indicate a lower reported exposure, and values equal to 1.0 indicate no meaningful change. These values may reflect heterogeneous study designs, dosing regimens, and patient populations. It is noted that regulatory authorities (FDA) define the intensity of interactions by the fold-change in AUC: a strong inhibitor causes a more than fivefold increase in AUC, a moderate inhibitor - 2-5 times, and a weak inhibitor - 1.25-2 times [10], [21]. This highlights the usefulness of numerical thresholds for describing the severity of drug interactions.

*2.2. Drug Name Normalization and Dataset Integration*

Integration of the two datasets required systematic normalization of drug nomenclature to enable accurate matching across different naming conventions. To reconcile naming conventions (generic/brand/variants), we implemented a multi-step matching pipeline: (1) exact matching of standardized generic names; (2) fuzzy matching (Levenshtein threshold 0.9); (3) synonym expansion using DrugBank synonyms; (4) targeted manual curation for high-frequency entities. This procedure matched 86.4% of PK-DDIP drug names to DrugBank entries, yielding 1,803 unified interaction pairs.

Directionality policy: PK-DDIP interactions are directional (perpetrator > victim). DrugBank pairs are treated as undirected text records (A–B). During integration we join on {A,B} irrespective of order, but we preserve the PK-DDIP direction when reporting AUC effects (i.e., A increases/decreases the exposure of B). We do not infer direction from DrugBank text; when only undirected text is available, we present qualitative context without a quantitative arrow.

*2.3. Web Application Development*

The web application was built with Python technologies for data-intensive scientific workflows (Fig.1) [22]. The frontend is implemented in Streamlit [23], enabling rapid development of interactive data applications. Data manipulation and analysis use pandas (v2.3), leveraging efficient tabular data structures and vectorized operations [24].

A cascading search constrains the second drug list to valid partners for the first, reducing user actions and preventing invalid combinations.

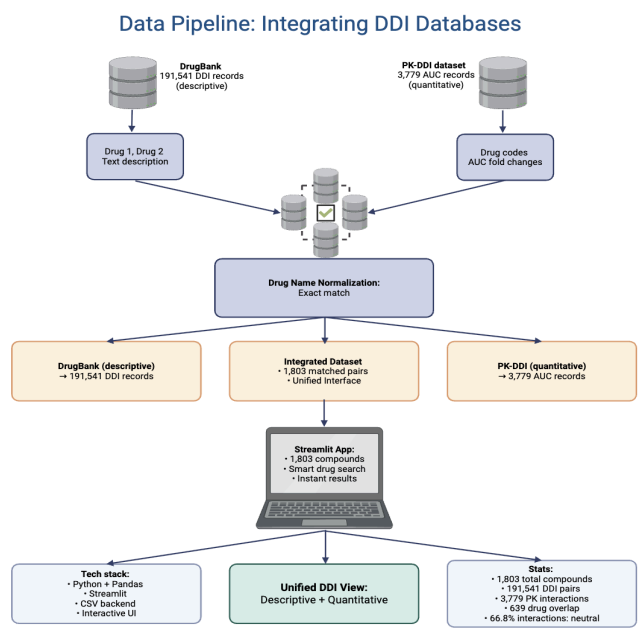

**Fig. 1.** Data Pipeline

## 3. Results

*3.1. Dataset Characteristics and Descriptive Statistics*

The DrugBank-derived set contains 191,541 interaction records among 1,701 distinct compounds. Cardiovascular agents (18,247 interactions, 9.5%), CNS medications (16,892 interactions, 8.8%), and anti-infectives (14,738 interactions, 7.7%) are prominent. Approximately 67% of records include specific mechanistic information; 33% provide broader pharmacodynamic descriptions.

The PK-DDIP dataset comprised 3,779 unique drug interaction pairs with AUC fold-change values extracted from FDA-approved labeling (DailyMed), involving 741 distinct compounds [5]. The dataset covered 23 major therapeutic categories, with the highest representation in cardiovascular medications (892 interactions, 23.6%), immunosuppressants (445 interactions, 11.8%), and psychiatric medications (387 interactions, 10.2%). AUC fold-change values ranged from 0.12 to 28.4, with a median value of 1.23 (interquartile range: 0.89-2.15).

*3.2. AUC Fold-Change Distribution Analysis*

Across 3,779 PK-DDIP records, distribution by category was: increased exposure (>1.0) 33.9% (1,281), decreased (<1.0) 19.8% (750), no change (=1.0) 46.3% (1,748). By magnitude: strong (>5.0 or <0.2) 3.4% (128), moderate (2.0–5.0 or 0.2–0.5) 9.3% (353), weak (1.0–2.0 or 0.5–1.0) 41.0% (1,550).

Several interactions exhibited extreme AUC changes with substantial clinical implications. Examples with large changes (are computed directly from PK-DDIP label-derived AUC ratios) include ritonavir–fluticasone (~350-fold), fluvoxamine–ramelteon (190-fold), cyclosporine–rifaximin (124-fold), nefazodone–buspirone (50-fold) [25], [26]. On the reduction side, rifampicin with isavuconazonium (0.003), and rifampicin with voriconazole or midazolam (0.040) illustrate substantial decreases.

### 3.3. Data Coverage and Integration Analysis

We found 1,667 PK-DDIP pairs (53.9%) with corresponding descriptive interactions in DrugBank, enabling unified qualitative + quantitative presentation. Conversely, 189,468 DrugBank pairs (99.1%) lacked corresponding AUC values, underscoring the qualitative–quantitative gap.

ATC mapping used DrugBank-provided ATC codes. ATC mapping indicated broad coverage: alimentary tract & metabolism accounted for 51.7% of drugs; cardiovascular 14.0%. In total, 86.0% mapped to at least one ATC class, and 19.3% to multiple classes [27].

### 3.4. Web Application Functionality

The web application interface consists of a streamlined, single-page design optimized for efficient DDI exploration. The primary interface features two sequential drug selection components positioned at the top of the page, followed by a results display area that dynamically updates based on user selections (Fig. 2). The first drug selection component presents the list of drugs; the second dropdown is filtered to valid interacting partners.

🍬 **Drug-Drug Interaction**

Select the first drug

| Cyclosporine |

Select the second drug

| Rifaximin |

[ Find ]

**Interaction Descriptio between *Cyclosporine* and *Rifaximin*:**

The serum concentration of **Cyclosporine** can be increased when it is combined with **Rifaximin**.

AUC of **Cyclosporine** is increased by **Rifaximin** (×124.00)

**Fig. 2.** Application interface

The search functionality implements intelligent matching algorithms that prioritize results based on search relevance, displaying drugs with names beginning with the search term first, followed by drugs containing the search term within their names.

Results display descriptive text with selective highlighting (drug names in bold; "increased/decreased" color-coded) and standardized AUC narratives indicating victim, perpetrator, and magnitude. The standardized AUC narrative mirrors regulatory conventions for reporting exposure changes [9], [10].

*3.5. Application Performance and Scalability*

In local testing, initial dataset loading completed within a few seconds; individual searches typically returned results in <1 s. The cascading design reduced user actions versus independent dual-selection. Query generation was instantaneous across tested pairs, with stable responsiveness across interaction complexities and result sizes.

## 4. Discussion

We present an open web tool that collates descriptive DDI text with reported AUC fold-changes from labeling, integrating qualitative context with quantitative effect size in one interface. Unifying >190k qualitative statements with 3,779 PK-labeled interactions addresses a persistent gap between narrative compendia and PK repositories [28].

The development of an interactive, web-based interface for integrated DDI assessment offers several advantages over traditional database consultation approaches. The streamlined UI lowers cognitive load, and standardized phrasing reduces misinterpretation risk. The visual highlighting of key interaction terms and color-coded presentation of quantitative effects enhance information processing and reduce the likelihood of misinterpreting interaction significance.

Limitations include partial quantitative coverage: only a subset of labeled interactions include explicit AUC values. The limited quantitative coverage particularly affects newer drug entities and less commonly prescribed medications, creating potential bias toward well-studied, established medications. PK-DDIP values derive from labeling and may reflect heterogeneous sources (dedicated clinical studies, population PK, modeling/simulation), with variability in design, dosing, populations, and analytics. Contextual modifiers (dose, route, timing, organ function, pharmacogenetics) are not yet modeled.

**Clinical use disclaimer.** This tool is a research prototype intended for education and hypothesis generation; it does not replace clinical judgment or institutionally validated CDS systems. Quantitative values originate from regulatory labeling and literature and may not reflect patient-specific factors (dose, route, organ function, pharmacogenetics). Do not use as a sole basis for prescribing decisions.

## 4. Conclusion

The platform offers fast access to harmonized qualitative and quantitative DDI information for researchers, educators, and clinicians. Future work will expand quantitative coverage through literature mining and additional repositories, incorporating uncertainty annotations. The incorporation of machine learning approaches for DDI prediction based on chemical structures and pharmacokinetic properties holds significant promise for addressing current coverage limitations while providing uncertainty estimates for clinical interpretation.

## Acknowledgments

The authors thank the teams behind DrugBank and the PK-DDIP project for providing access to their comprehensive interaction datasets, which made this work possible. We also acknowledge the open-source community for developing the Streamlit framework and the pandas library.

## Data & Code availability

Code (app + integration scripts) is available at: [https://github.com/diadkinang/drug-drug-interaction]. Version archived as v1.0.0 (Git tag); dependencies pinned in requirements.txt. PK-DDIP and DrugBank data were accessed under their respective licenses; we do not redistribute DrugBank interaction text in bulk.